\documentstyle[12pt]{article}
\begin{document}

\title{On the Quantum Baker's Map and its Unusual Traces}
\date{}
\author{Arul Lakshminarayan  \\
{\sl Physical Research Laboratory,}\\
{\sl Navarangapura, Ahmedabad,  380009, India.}}
\maketitle

\newcommand{\newc}{\newcommand}
\newc{\beq}{\begin{equation}}
\newc{\eeq}{\end{equation}}
\newc{\beqa}{\begin{eqnarray}}
\newc{\eeqa}{\end{eqnarray}}
\newc{\pa}{\partial}
\newc{\bom}{\boldmath}
\newc{\btd}{\bigtriangledown}
\newc{\rarrow}{\rightarrow}
\newc{\ep}{\epsilon}
\newc{\beqas}{\begin{eqnarray*}}
\newc{\eeqas}{\end{eqnarray*}}
\newc{\mb}{\mbox}
\newc{\tm}{\times}
\newc{\hu}{\hat{u}}
\newc{\hv}{\hat{v}}
\newc{\hk}{\hat{K}}
\newc{\ld}{\lambda}
\newc{\sg}{\sigma}
\newc{\p}{\psi}
\newc{\kt}{\rangle}
\newc{\br}{\langle}
\newc{\ra}{\rightarrow}
\newc{\dg}{\dagger}
\newc{\non}{\nonumber}
\newc{\ul}{\underline}
\newc{\longra}{\longrightarrow}
\newc{\hs}{\hspace}
\newc{\eps}{\epsilon}
\newc{\longla}{\longleftarrow}
\newc{\ts}{\textstyle}
\newc{\f}{\frac}
\newc{\ol}{\overline}

\begin{abstract}
The quantum baker's map is the quantization of a simple
classically chaotic system, and has many generic features that
have been studied over the last few years. While there exists a
semiclassical theory of this map, a more
rigorous study of the same revealed some unexpected
features which indicated that correction terms of the order
of $ \log(\hbar)$ had to be included in the periodic orbit
sum. Such singular semiclassical behaviour was also found in
the simplest traces of the quantum map. In this note we
study the quantum mechanics of a baker's map which is obtained
by {\it reflecting} the classical map about its edges, in an
effort to understand and circumvent these anomalies.
This leads to a {\it real} quantum map with traces that follow
the usual Gutzwiller-Tabor like semiclassical formulae.
We develop the relevant semiclassical periodic orbit sum for this
map which is closely related to that of the usual baker's map,
with the important difference that the propagators leading to
this sum have no anomalous traces.

\end{abstract}
\vspace {2in}
\begin{center}
{\bf To Appear in Annals of Physics.}
\end{center}
\newpage
\section{Introduction}

\hspace{.5cm} The semiclassical analysis of classically non-integrable systems
has been an active field of research in the past few years. One
of the central problems is the determination of the spectra of
classically chaotic systems. To date the only approach to this
has been via the semiclassical periodic orbit sum, which relates
the traces of quantum propagators to a weighted sum over classical
 periodic orbits. The classical ingredients are the
actions of the periodic orbits, their stabilities and their Maslov
indices [20]. While this sum has been the backbone of much of
the semiclassical analysis of classically chaotic systems it is
known to be plagued by a lack of convergence and a loss of
accuracy with increasing time. Thus the study of simple systems
like certain piecewise linear maps on the torus can be very
helpful in understanding some of the difficulties involved.
Simple  abstract maps have been quantized and studied over the years,
including the well known family of continuous automorphisms, the
 cat maps [9,14,15,16], and the piecewise linear baker's map [9,4,5].
Unfortunately the cat maps have a very non-generic
quantum mechanics, with an exactly periodic propagator (in
time), and a periodic orbit sum that is {\it exact}.

The quantum baker's map [4,5] has  become a textbook example of
quantizing a simple classically chaotic system [6]. The
simplicity of the classical map is preserved in the quantization
which has revealed many generic features, including random matrix
like spectral fluctuations and eigenfunction scarring [4,5,7].
In the semiclassical regime a periodic orbit sum can be written
for the traces of the propagator that is formally like the Gutzwiller-
Tabor periodic orbit sum [2,3]. However all such
analysis treat the fixed points, and hence the simplest
trace of the map, separately as there is no linearizable neighbourhood
about such points. A more careful semiclassical analysis [1]
 which paid attention to the essential discreteness of the
quantum map revealed that there were terms of the order of $\log
(\hbar)$ in the simplest traces of the map and hence such
corrections will have to be incorporated into the periodic orbit
sum. These semiclassically divergent terms deserve to be further
explored to see if they are generic of classically chaotic
systems allowing a Markov partition or whether they are peculiar
to the quantum baker's map.

The baker's map
is essentially defined on a square with boundaries. Imposing
 periodicity, classically, makes the fixed points non-generic,
in the sense that the corner fixed points at $(0,0)$ and $(1,1)$
become ``half hyperbolic points''
and could give rise to unusual semiclassical behaviour. Thus we
would ideally like to impose some other boundary conditions
while quantizing the map.
Here we will adopt reflective boundary conditions on the {\it classical}
map itself, thereby making it a ``redundant'' four bakers' map.
{\it This will create
four classically disjoint baker's maps}. Now this map will by its
very construction have a natural topology on a torus. A
quantization of this map is then possible along the  lines
adopted in the original quantization of the baker's map
[4]. We will find that the trace of this quantum map has
{\it no semiclassically divergent terms}. We also will prove
that the trace of the time $T$ semiquantum propagator, obtained by
quantizing the classical $T$ step map has no $\log(\hbar)$ terms,
semiclassically.

	In section 2 we will briefly review the classical baker's
 map [9,5] and discuss its 4-fold (classically redundant) version. In
section 3 we will write down the quantum map after discussing
some essential features of quantum mechanics on a torus. In
section 4 we discuss the semiclassical trace of the redundant baker's
map
and compare it with the trace of the usual quantum baker's map.
We compare the semiclassics of the semiquantum at time two for
the two maps.
We will also study numerically the convergence of the traces of
the exact and the semiquantum propagators at time two. In
section 5 we will generalize some results of section 4 and
develop the semiclassical periodic orbit sum which has the canonical
form of the Gutzwiller-Tabor sums.  In section 6 we end with
a discussion of  our results.

\section{The Classical Maps}

	\hspace{.5cm} The classical baker's map is given by :

\beq
       ( q ^{ \prime},p^{\prime})\, = \,
 \left\{ \begin{array}{ll}
  (2\,q,\,p/2) & \mbox{if} \; \; 0 \leq q < 1/2  \\
          (2\,q -1, \, (p+1)/2 ) & \mbox{if} \; \;  1/2 \leq q < 1.
        \end{array}
	\right.
\eeq
$q$ and $p$ are interpreted as ``position'' and ``momentum'' in
dimensionless units. The phase space is a half open square and hence
toral boundary conditions may appear  natural. There is
one fixed point at the origin $(0,0)$. If we tessellate the plane
with such open squares it is equivalent to having imposed toral
boundary conditions.  When we examine the behaviour of points
in the neighbourhood of the fixed point we find  that
they do not have a generic hyperbolic structure, in that they
are only a ``half of a hyperbolic point''. An unusual classical
picture emerges if we impose periodic boundary conditions in
position and momentum.  This also becomes
clear if we define the map on the closed square instead of the
half open one. Then there are two fixed points $(0,0)$ and $(1,1)$,
while the points $(0,1)$ and $(1,0)$ are not fixed. Thus
we cannot treat the four corners of the classical map on a same
footing as two of them are fixed points with a half neighbourhood
of a normal hyperbolic point while the other two are not fixed.

 The dynamics of the area preserving
map eq.(1) is well known.  It is completely chaotic, has a Lyapunov exponent
of $\log 2$, and is a Bernoulli shift on two symbols. Further details
of this map may be found in the references. Quantization of this
map was first done in [4] wherein periodic boundary conditions
were imposed on the states. While this is the most elegant
quantization, as has been noted above, it has some unusual
semiclassical features.  While the quantum properties ``in the bulk''
are properly accounted for, there are logarithmically divergent ``corner
regimes'' [1], creating significant departures from
standard semiclassical results. Starting from the notion that these are
generated by the boundary conditions of the quantum map we seek
to modify the classical map itself with the least damage to its
dynamics.

We first impose reflective boundary conditions on the classical
baker's map. Imagine  that the edges of the baker defined
on the closed square are mirrors. After one reflection
about the edges it is clear that the images tessellate the
plane and we can impose toral boundary conditions (figure 1).
We can do this for any map on the square. We consequently
have four baker's maps that are disjoint and are on a torus.
Thus the classical map is not globally ergodic, but has four distinct
regions of equal area where the dynamics is chaotic.

We can scale the map so that the four bakers are on the unit
square. This will be the map that we will quantize. It is
clear that classically this map is just the baker's map
that is made three times redundant, but quantum mechanically
there are essential differences. Most importantly the four
baker's maps can now be on a torus. The four corners of the unit
square are fixed points and so is the center $(1/2,1/2)$.
These are generic fixed points.{\it Also while the four classical
baker's are isolated, quantum mechanically they are not}.
There will be {\it tunnelling} amongst these classically
non-interacting bakers. Such tunnelling effects in the context of
other juxtaposed bakers' maps are also discussed in [10].

The extremely piecewise nature of the four
bakers' map may appear to preclude quantization, as one
cannot define quantum projection operators on these
partitions. In fact we can view the classical map more
``holistically'', and this will be the key to quantization.
We can write the four bakers' map as follows.

\beq
       ( q ^{ \prime},p^{\prime})\, = \,
 \left\{ \begin{array}{ll}
  (2\,q(\mbox{mod} 1),\,(p+[2p])/2) & \mbox{if} \; \; 0 \leq q < 1/4  \\
	&                   \mbox{ or if} \; \; 3/4 \leq q <  1 \\
          (2\,q -1/2, \, (p+1/2)/2 ) & \mbox{if} \; \;  1/4 \leq q < 3/4.
        \end{array}
	\right.
\eeq

Square brackets stand for the integer value function. Thus we
see that in one time step the rectangle $(1/4,3/4) \times (0,1)$
gets linearly transformed into $ (0,1) \times (1/4,3/4)$ (figure
1). We can
think of the dynamics in the other partitions similarly. First
shift all the partitions 1,2,3,4 by (1/2,1/2), so that the
fixed point at the corner gets shifted to the center.  One
can then perform the same operation as for the partitions 5,6,7,8,
and finally shift the fixed point at the center back to the corner.
This is exactly how we will quantize the map, as it then
requires dividing the Hilbert space into two orthogonal subregions,
one corresponding to the classical partitions 5,6,7,8 and the
other to the partitions 1,2,3,4, both of which trivially exist.

\section{Quantum Map}

\subsection{Preliminaries}

\hspace{.5cm} We will now discuss some aspects of quantum kinematics on the
torus [8,5] that are essential for the quantization of the four  bakers'
 map.
Imposing periodic boundary conditions on both position and momentum,
results in a state space of finite dimensions. The Planck constant is
related to the dimensionality of the space $N$ by \[2 \pi \hbar \;= \;
N^{-1} .\] The quantization in [4] was modified in [5],
incorporating antiperiodic boundary conditions on the states to
restore the full symmetry of the classical map, and this we will adopt.

The position and momentum eigenvectors are given by
\[    |q_{n} \kt =\;\;| \frac{n+1/2}{N} (\mbox{mod}1) \kt \;\; n=0,1,...,N-1,\]
  \[    |p_{m} \kt =\;\;| \frac{m+1/2}{N} (\mbox{mod}1) \kt \;\; m=0,1,..,N-1,
\]
with the transformation functions being discrete versions of
plane waves:
\beq \br p_{m}|q_{n} \kt = N^{-1/2}e^{-2\pi i(m+1/2)(n+1/2)/N} \equiv
(G_{N})_{mn}.
\eeq

$G_{N}$ is  a discrete Fourier transform on $N$ sites.
The symmetry operation of reflection about the
center of the square $(q \, \ra  \, 1-q,\, p\, \ra \, 1-p)$ is
incorporated quantally by the matrix $R_{N}$ defined as
\beq \br q_{n}|R_{N}|q_{n^{\prime}} \kt = \delta (n + n^{\prime}
+1,  N)
\eeq
which is zero except for the secondary diagonal which has ones.
The position translation operator is $V$. It is such that
\beq \br q_{n} |V = \br q_{n+1} |, \;\;\mbox{and}\;\; V^{N}=-1.
\eeq
The translation operator in momentum, $U$,  is similarly defined.
The analogue of  the uncertainty relation is the
non-commutativity of $U$ and $V$ described by
\beq UV \, = \, e^{-2\pi i /N} VU. \eeq
The matrix $V$ is diagonal in the momentum basis,
\beq
\br p_{m}|V| p_{m^{\prime}} \kt = \delta_{m,m^{\prime}} e^{2 \pi
i (m+1/2)/N}
\eeq
and similarly $U$ is diagonal in the position  basis.

There are several useful relations amongst the matrices
introduced above and all of them may be verified by direct computations.
Further identities in [5] are also useful in proving
some of the relations below. The act of classically
translating by (1/2,1/2) on the torus is an essential ingredient
of the quantization. Hence we will define the unitary operator [5]
\beq
	T \, =\, e^{i \pi N/4} U^{N/2} V^{-N/2}.
\eeq
We note that as long as $N$ is divisible by four (which we
assume throughout this paper) the operators $U^{N/2}$ and $V^{-N/2}$
commute, thus there is no ambiguity in defining this operator.

$T$ has the following properties;
\beq
	\begin{array}{rcl}
	T^{2}&=&1, \\
	 T\, G_{N} & =& - \, G_{N} \, T,
	\end{array}
\eeq
and
\beq
	T\, R_{N} \, =\, R_{N} \, T,
\eeq
which may be established by using the identities given above and
those in [5].

\subsection{Quantum dynamics}

With the preliminaries established above, one may quickly go to
the heart of the quantization [4] by first quantizing the transformation
of the partitions 5,6,7,8 (figure 1). In the {\it mixed}, momentum-position,
representation it is given by the matrix operator which we
denote as $u_{0}$, where

\beq
u_{0} = \left( \begin{array}{cccc}
			0 &0 & 0&0 \\
			0 &g_{11}&g_{12}&0 \\
			0 & g_{21} &g_{22}& 0    \\
			0 &0&0&0
			\end{array}
			\right ).
\eeq
and
\beq
G_{N/2} \equiv \left( \begin{array}{cc}
			g_{11}&g_{12}\\
			g_{21}&g_{22}
			\end{array}
			\right).
\eeq

$ G_{N/2} $ is the $ N/2 $ dimensional Fourier transform that is
centrally placed in $ u_{0} $ which is itself an $ N $ dimensional
matrix. The $ N/4 $ dimensional matrices $ g_{ij} $ forming the larger
matrix $ G_{N/2} $ are written here for future usage.
The rest of the matrix elements in $u_{0}$ are zero. Thus in
position representation the transformation of the partitions
5,6,7,8 is given by the matrix $ G^{-1}_{N} \, u_{0} $.

Following the discussion of the classical map in the previous
section, we can now write the transformation of the partitions
1,2,3,4 as
\beq
T^{-1} \, G_{N}^{-1} \,u_{0} \, T \;= \; T \, G_{N}^{-1}\,u_{0}\,T,
\eeq
utilizing the identity $ T^{2}=1 $.
Thus the whole quantum map, in the position representation, for the
four bakers' map is given by the $N$ dimensional matrix
\beq
	 \begin{array}{lcl}
	B & =& T \, G_{N}^{-1} u_{0} T \, +\, G_{N}^{-1}
u_{0} \\
	& =& G_{N}^{-1}(\, -T u_{0} T \, + \, u_{0}).
	\end{array}
\eeq
In the last equality we have used the anticommutation property
in eq.(9).  We have added the two matrices for the
transformation of the partitions 1,2,3,4 and 5,6,7,8 as the
  corresponding quantum subspaces in
the Hilbert space are mutually orthogonal. Though
 unitarity is not manifest, there are certain
advantages of this representation of the quantum map that we can
immediately exploit.

The advertised advantages are the symmetry properties of the
quantum map. The classical map has the symmetries of reflection
about the center of the square and translation in phase space by
$(1/2,1/2)$. The corresponding quantum symmetries are implemented
by the matrices $R_{N}$ and $T$, eqs.(4,8). From the
representation of the quantum map $B$ in eq.(14), it is clear
that
\beq
[B,\, T]\, =\, 0, \;\; [B, \, R_{N}]\, =\, 0.
\eeq
The square brackets define the usual commutator, and the above
follows from the commutation of $T$ with $R_{N}$, eq.(10), and
the commutation of $R_{N}$ with $ u_{0} $. Thus the quantum map
has two exact symmetries corresponding to the classical map.
Since $T^{2}=1,\,R_{N}^{2}=1$ the eigenvalues of the operators
$T$ and $R_{N}$
are $ \pm 1 $, hence the eigenfunctions can be labelled by
two good quantum numbers, and there are four symmetry classes.

There is another classical symmetry that is obvious from the way
we have constructed the four bakers' map, the symmetry of
reflection about the line $q=1/2$ (and $p=1/2$). This is translated quantally
by requiring that the quantum map $B$ satisfy the relation
\beq
	R_{N}\, B^{\ast}\, R_{N} \, =\, B.
\eeq
Here $\ast$ denotes complex conjugation.
This symmetry is an antiunitary symmetry and is hence
implemented by  $R_{N} \, K $, where $K$ is the
complex conjugation operator. When this symmetry is combined with the
commutation of $B$ with $R_{N}$, eq.(15), we get the
condition
\beq
	B^{\ast}\, =\, B,
\eeq
that is $B$ is a {\it real } quantum map. Thus the unitarity of $B$
will infact establish it to be an {\it orthogonal} matrix, and
the quantum four bakers' map can also be thought of as a ``simple''
irrational rotation in a real $N$ dimensional vector space.

Another antiunitary symmetry of the map, which it shares with the
usual quantum baker's map, is the symmetry of time reversal [4].
The classical symmetry is one of interchanging position and
momentum and going one step back in time. This is implemented
by the operator $G_{N} \; K$, where $K$ is once more the
conjugation operator. Thus we require the quantum map $B$  to satisfy
the condition that
\beq
	\begin{array}{lcl}
	G_{N}\, B^{\ast}\, G_{N}^{-1}&=& B^{-1}\;\;\mbox{or}\\
	G_{N}\,B \, G_{N}^{-1}&=& B^{t},
	\end{array}
\eeq
	where we have used the as yet unexplicit reality
and orthogonality of $B$,
and $B^{t}$ is the transpose of $B$. The cheapest way to verify
the above is to take the transpose of $B$ as defined in eq.(14),
and noting that while $G_{N}^{-1}$ and $u_{0}$ are symmetric,
$T$ is antisymmetric. Thus the quantum four
bakers' map is time reversal symmetric.

We will now write the quantum map $B$ in more explicit forms
so that its reality and orthogonality will become evident.
Using the definition of the translation operator $T$ in eq.(8),
we can write $B$ from eq.(14) as the matrix
\beq
	B\,=\,G_{N}^{-1} \left( \begin{array}{cccc}
		(-1)^{N/4}g_{11}&0&0&(-1)^{N/4}g_{12}\\
		0 & g_{11} & g_{12} & 0 \\
		0 & g_{21} & g_{22} & 0 \\
		(-1)^{N/4}g_{21} & 0&0& (-1)^{N/4}g_{22}
		\end{array}
		\right).
\eeq
(For some more details see the Appendix). From this form one may
infer the unitarity of $B$, as $G_{N}^{-1}$
is an unitary matrix, and the second matrix product's unitarity
follows from that of $G_{N/2}$. We note here some features
of this matrix. Firstly there is a Planck constant dependent
phase in front of one block of the Fourier transform that
corresponds to the classical transform of the partitions
1,2,3,4. This phase factor is simply $ \pm 1$  as we have assumed
$N$ to be divisible by four. But for this phase factor, $T$
symmetry would be exact only if $N$ is divisible by 8. Secondly
this matrix form might have been written by inspection using the
methods outlined in [4]. While this would have missed out
the phase factor, we note that such phase factors sitting
globally on blocks do not affect the classical limit. They might
affect such properties as traces, and symmetries.

Further evaluation of matrix elements is possible, using the
basic definitions of the components of $B$ as Fourier
transforms. After performing a few elementary geometric sums,
we get the following explicit real representation of $B$ in the position
basis, $\br q_{n^{\prime}}|B|q_{n} \kt $, as
\beqa
 &&(-1)^{N/4} \frac{\sqrt{2}}{N}\, \frac{\sin \left[ \pi(n^{\prime}-2\,
n-1/2)/2 \right]}{\sin \left[ \pi(n^{\prime}-2\, n-1/2)/N \right]} \; \;\;\;
\mbox{if}\;\;\;\; 0 \leq n
\leq N/4-1, \nonumber \\
 &&\frac{\sqrt{2}}{N}\, \frac{-\sin \left[ \pi(n^{\prime}
+1/2)/2 \right]}{\cos \left[ \pi(n^{\prime}-2\, n-1/2)/N \right]}\;\;\;\;
 \mbox{if} \;\;\;\; N/4 \leq n
\leq 3N/4-1,  \\
 &&(-1)^{N/4} \frac{\sqrt{2}}{N}\, \frac{-\sin \left[ \pi(n^{\prime}-2\,
n-1/2)/2 \right]}{\sin \left[ \pi(n^{\prime}-2\, n-1/2)/N \right]} \;\;\;\;
 \mbox{if} \;\;\;\; 3N/4 \leq n
\leq N-1. \nonumber
\eeqa
Due to the time reversal symmetry, the momentum space
representation is also real, and infact just the transpose of
$B$. The matrix elements of the quantum map are thus simple
elementary functions, a feature that it shares with the
single baker's map. The classical limit of $B$ corresponds
to the four fold redundant baker, and this occurs in the same
way as for the single baker's map. From the classical map it
may be expected that the quantum map be nearly block diagonal
(with blocks of size $N/2$) in {\it both} position and momentum
basis. This is infact the case, and moreover the ``tunnelling''
matrix elements are not exponentially small in Planck constant,
but rather go as $1/N$ as may be inferred from eq.(20). This
leads to rather large tunnelling effects as time goes by.

\subsection{Properties of the Quantum Map}

\hspace{.5cm} The quantum four bakers' map is real and preserves all
classical symmetries. It is therefore of interest to study the
spectral properties of this map, and compare it with those of
the quantum baker's map. As this takes us away from the central
thrust of this article we will merely state some results,
although a much more comprehensive study is definitely
desirable. The symmetry group underlying the quantum map is
isomorphic to the
Abelian dihedral group $D_{2}$ consisting of four elements
$(I,\; R,\;T, \; RT)$. Here $I$ is the identity matrix.
The symmetry $R\,T$, the combined symmetry of reflection
about the center of the square and translation in phase space
by 1/2 is the ``$R$ symmetry'' of the usual baker's map, which
corresponds to the symmetry of reflection about the center of
each {\it individual} baker's map. It is satisfying that even
this symmetry is present in the quantum map.
 The existence of the discrete commuting
 canonical symmetries
of $R$ and $T$ makes the labelling of states by two
quantum numbers possible. In particular the eigenstates belong
to  one of the four possible classes, denoted as $|+ \, + \kt ,\;
| + \, - \kt , \; | - \, + \kt ,  \; | - \, - \kt $. All the
eigenangles (in units of $2 \pi$) are irrational, a property it
shares with the quantum baker's map.

There are no degeneracies in the quantum map. However the reality of the
map  implies that eigenvalues occur as complex conjugate
pairs, which belong to {\it different}  irreducible representations
of the quantum map (with different $T$ eigenvalues), and hence to
different symmetry classes.
The reduced four bakers' map is block diagonal with four blocks.
The following unitary matrix simultaneously diagonalizes both
$R$ and $T$, and hence can be used in the reduction of the
quantum map. If we write
\[ P\;=\;\]
\beq
 \frac{1}{2}\left( \begin{array}{cccc}
	1&1&1&1 \\
	-i\,R_{N/4}\,A_{N/4} & i\,R_{N/4}\,A_{N/4} & i\,R_{N/4}\,A_{N/4}
 & -i\,R_{N/4}\,A_{N/4}\\
	-i\,A_{N/4} & i\,A_{N/4} & -i\,A_{N/4} & i\,A_{N/4} \\
	R_{N/4}&R_{N/4}&-R_{N/4}&-R_{N/4}
	\end{array}
	\right),
\eeq
then $P^{-1}\,B\,P$ is a block diagonal matrix with four blocks.
Where
\[ (A_{N/4})_{mn} \,=\, (-1)^{m} \delta_{mn},\;\;\;m,n\,=\, 0,
\ldots, N/4-1.\]
One may exploit these symmetries to reduce the size of matrices
that need diagonalization. For example
all $ | + \, + \kt $ eigenvalues can be got by diagonalising the
$N/4$ dimensional matrix
\[  B_{11} - i B_{12}R_{N/4} A_{N/4}  -i
B_{13} A_{N/4} + B_{14} R_{N/4}, \]
  where $B_{mn}$ is the $mn^{\mbox{th}}$
block of the quantum map
divided into sixteen $\;\;N/4$ dimensional matrices.

\section{The Semiclassical Traces}

\subsection{The Anomalous Traces}

\hspace{.5cm} In this subsection we recall some features of the quantum baker's
map and its unusual traces [1].
 Consider the
quantum baker's map,  denoted here as $B^{\prime}$ and defined as
\beq
	B^{\prime} \,=\, G_{N}^{-1} \left( \begin{array}{cc}
				G_{N/2}& 0\\
				0    & G_{N/2}
				\end{array}
				\right).
\eeq
The simplest trace in the quantum problem is Tr$(B^{\prime})$,
and in [1], we find its semiclassical (large N) form,
\beq
\mbox{Tr} B^{\prime} \sim 2 \sqrt{2} \left[ \frac{1}{4} -
\frac{i}{2 \pi} \left( \log (8 N/\pi) + \gamma \right) \right ],
\eeq
where $\gamma$ is the Euler constant.

An aspect of the quantum baker's map that has been exploited in its
semiclassical analysis is the possibility of quantizing the map
after a certain number of classical iterations. Thus for
instance if we quantize the baker's map after two time steps
($B^{(2) \prime})$,
we would not get the same operator as $B^{\prime 2}$. The
differences between the two families of operators (quantum and
semiquantum, in the terminology of ref. [1]) is of great
interest, as it reflects the differences between quantal and
classical evolutions. In [1] we find a careful
semiclassical comparison of the matrix elements of the above two
operators, and beautiful visual representations of the same.

Part of the differences may be captured in the trace metric [11],
 as we can derive the following semiclassical result;
\beq
 \mbox{Tr}(B^{\prime 2}B^{(2) \prime \dagger})\,\sim \,
N-\frac{4}{\pi ^{2}}\log (N) + \alpha,
\eeq
where $ \alpha = \frac{4}{\pi ^{2}}(\log (\frac{\pi}{8})+4 \beta (2)
- \gamma -1) \approx 0.46686.... $
$\beta (2)$ is Catalan's constant [12].
The singular semiclassical term in both of the above traces
is logarithmic. The trace of the propagator is governed semiclassically
by the fixed points of the map, and the logarithmic  behaviour
in the first trace is
a reflection of the corner configuration of this fixed point.
Eq.(24) measures the ``distance'' between the two operator families,
the quantum and the semiquantum in the Euclidean norm, and it seems
to indicate semiclassical divergence (in the norm) even if only marginally.

The trace of the propagator at time two is itself difficult to
estimate. While we can do a semiclassical evaluation of the trace
of $B^{(2) \prime }$, the same for $B^{\prime 2}$ is not available, except for
some partial results in [1].  In ref. [1]
the semiclassical form of Tr$(B^{(2) \prime})$ is decomposed in accordance
with classical symbol sequences. This exact symbolic decomposition
of the propagator at {\it both} the quantum and the semiquantum
level is an interesting feature of the baker's map, and is true
only at the semiquantum level for the four bakers' map. Let the
quantum propagator associated with the transformation of the
partition 00 be $b^{\prime}_{00}$. Partition 00 is the first
of the four equal  vertical partitions of the phase space square.
 For complete definitions see ref. [1].

For our immediate purposes it is sufficient to
note that the  trace of $b^{\prime}_{00}$ is governed semiclassically
by  the fixed point (0,0). Similarly the transformation of the
partition 10 is given by the propagator $b^{\prime}_{10}$, whose trace
is semiclassically governed by the fixed point (1/3,2/3). Note
that for the time two map, there are four classical fixed
points, and that the propagators $b^{\prime}_{00,10}$ are not
unitary. The  complete trace of $B^{(2) \prime}$ is given by
$2 \; \mbox{Tr} (b^{\prime}_{00} \;+\; b^{\prime}_{10})$. This is a
consequence of $R$ symmetry [1]. Again in ref. [1] we find the
semiclassical form of these traces which we repeat here for
comparison with the corresponding four bakers' traces.
\beq
\mbox{Tr} (b^{\prime}_{00}) \; \sim \; \frac{-i}{3 \pi} \log\; N \;+\;
	\frac{2}{3} \left\{ \frac{1}{4} \; -\; \frac{i}{2 \pi}
\left[ \log \; \frac{8}{3 \pi} \;+\; 2 \log(1+\sqrt{2}) + \gamma
 \right] \right\} ,
\eeq
\beq
\mbox{Tr} (b^{\prime}_{10}) \; \sim \; \frac{2}{3} e^{4 \pi i N/3}
\eeq
We note that once again there are terms logarithmic in the
Planck constant in the semiquantum trace governed by the corner
fixed point. There  are no such terms for the trace of $b^{\prime}
_{10}$ which is controlled by the ``good'' fixed point in the interior.

\subsection{ Four Bakers' Semiclassical Traces}

\hspace{.5cm}We now evaluate the semiclassical trace of the four bakers' map,
whose real position basis representation is given by eq.(20).
Following ref. [1] we can do a careful aymptotic analysis, in order to avoid
possible pitfalls, but in fact elaborate analysis is not
necessary and we will see, as can be expected, that it is legal
here to replace summations with integrals. We will write the
traces in forms that explicitly resemble those of the usual
quantum baker's map. From the original definition of $B$ in eq.(19),
we will define the following non-unitary  matrices.
\beq
	b_{0} \;=\;e^{i \pi N/4} G^{-1}_{N} \left( \begin{array}{cccc}
	g_{11} &0&0&0\\
	0&0&0&0\\
	0&0&0&0\\
	0&0&0&0
	\end{array} \right)
\eeq
\beq
	\ol{b}_{0} \;=\;e^{i \pi N/4} G^{-1}_{N} \left( \begin{array}{cccc}
	0&0&0&0\\
	0&0&0&0\\
	0&0&0&0\\
	g_{21}&0&0&0
	\end{array} \right)
\eeq

Then we get
\beq
\mbox{Tr} \; b_{0} \; = \; e^{i \pi N/4}\,\frac{\sqrt{2}}{N} \;
\sum_{m=0}^{N/4-1}\,
\sum_{n=0}^{N/4-1} \; e^{-2 \pi i (m+1/2)(n+1/2)/N}
\eeq
and
\beq
\mbox{Tr} \; \ol{b}_{0} \; = \;e^{i \pi N/4}\, \frac{\sqrt{2}}{N} \;
\sum_{m=0}^{N/4-1}\,
\sum_{n=0}^{N/4-1} \; e^{2 \pi i (m+1/2)(n+1/2)/N}.
\eeq
These further imply that
\[
\mbox{Tr} \; (b_{0}\,+\,\ol{b}_{0})\;=\; e^{i \pi N/4} \,\frac{2 \sqrt{2}}{N}
\, \mbox{Re}\, \left( \sum_{m=0}^{N/4-1} \,\sum_{n=0}^{N/4-1}
\; e^{-2 \pi i (m+1/2)(n+1/2)/N} \right)
\]
\beq
=\;e^{i \pi N/4}\, \frac{2 \sqrt{2}}{N}.\frac{1}{4} \; \sum_{m=-N/4}^{N/4-1}\,
\sum_{n=-N/4}^{
N/4-1} \; e^{-2 \pi i (m+1/2)(n+1/2)};
\eeq
we get the final form  by reflecting the summation lattice about
$m=0$ and $n=0$.  This form belongs to the ``regular regime'',
that is there are no corners in the summation and asymptotically
it can be approximated by a  continuum integral, followed by a
stationary phase approximation. We then extend the range of
integration and evaluate the integral exactly, as it is a Gaussian.

 Thus we get
\beq
\mbox{Tr}\, (b_{0}\,+\, \ol{b}_{0}) \; \sim \;
e^{i \pi N/4}\,\frac{\sqrt{2}}{2 \, N} \; \int_{-\infty}^{\infty} dp \,
\int_{-\infty}
^{\infty}dp \; e^{-2 \pi i N pq} = e^{i \pi N/4} \, \frac{\sqrt{2}}{2}.
\eeq
Finally using both $R$ and $RT$  symmetries we get
\beq
\mbox{Tr} \, B \; = \; 4 \,\mbox{Tr}\, (b_{0} \, +\,
\ol{b}_{0}) \; \sim \;
 e^{i \pi N/4}\,2 \; \sqrt{2}.
\eeq
Thus we see the close relationship between the {\it real} part of the
quantum baker's trace, eq.(23), and the trace of the four bakers' map,
and it comes as no surprise that {\it semiclassically the regular
part of the trace of the baker's map is  quarter  that of the four
bakers' map's trace}. We will observe in the following that this
is true also beyond time one.

The absolute value of the asymptotic trace is what one will
expect  with the semiclassical Gutzwiller-Tabor periodic orbit sum,
although we have not attempted to develop the semiclassics of
the map yet (we will do so in section 5).
 We can also compare
the above form, eq.(31),  with the analogous sum of the usual quantum
baker's map. We have from [1]
\beq
\mbox{Tr}\, B^{\prime} \;=\; \frac{2\sqrt{2}}{N} \,\sum_{m=0}^{N/2-1}
\, \sum_{n=0}^{N/2-1}
\; e^{-2 \pi i (m+1/2)(n+1/2)/N}.
\eeq

Admittedly, while the four bakers' map promises more ``generic''
behaviour than the baker's map, beyond the first trace  we have not
{\it proved} that such is the case. However some qualitative
arguments may be put
forward towards this end. Beyond time one, a very comprehensive
semiclassical analysis for the baker's map has been attempted
in [1] for time two. They found for instance that not
only were the time two quantum and semiquantum traces significantly
different for the corner regime that is controlled by the fixed
point at the corner, but so also were the ``bulk traces'' governed
by the ``good'' period two points.

A qualitative argument put forward to explain this [1]
is the presence of aliasing, when there are ``quasi-stationary''
points in the saddle point approximation. At time two, such
points converge on the origin which anyway has a singular
semiclassical behaviour, leading to  discrepancies even in
the ``bulk''. In the quantum four bakers' map, since the corner
regime is not singular, we may expect such aliasing effects
to create only ``benign'' behaviours. Thus the semiclassical
four baker does indeed promise to be  more ``generic''.

\subsection{The Time Two Semiquantum Map}

\hspace{.5cm} As has been noted earlier, a key ingredient in the semiclassical
quantum baker's map has been the semiquantum propagators. At
time two this is constructed out of the quantization of the time
two classical map. Indeed the simplicity of the classical
baker's map and the versatality of the quantization procedure makes
this possible. The author is not aware of any other map where
the construction of such (non-trivial) semiquantum propagators exists. This at
once makes the quantum baker's map special as well as valuable.
The construction of semiquantum propagators is also possible for
the four bakers' map.

	Here however we will adopt the  quantization
scheme in [4] neglecting phase factors at blocks and then
getting the phase factors by {\it imposing} $T$ symmetry. It has been
mentioned earlier that this is possible at the level of the
time one map itself. For arguments concerning the general case
see the Appendix.
 We can write the semiquantum map for time two as
the matrix
\[ B^{(2)} \,=\, \]
\beq
G^{-1}_{N} \, \left( \begin{array}{cccccccc}
		e^{i \pi N/8}g_{11}&0&0&0&0&0&0& e^{i \pi N/8}g_{12}\\
		0&0&g_{11}&0&0&g_{12}&0&0\\
		0&g_{11}&0&0&0&0&g_{12}&0\\
		0&0&0&g_{11}&g_{12}&0&0&0\\
		0&0&0&g_{21}&g_{22}&0&0&0\\
		0&g_{21}&0&0&0&0&g_{22}&0\\
		0&0&g_{21}&0&0&g_{22}&0&0\\
		e^{i \pi N/8}g_{21}&0&0&0&0&0&0& e^{i \pi N/8}g_{22}
		\end{array}
		\right),
\eeq
where
\beq
		G_{N/4} \, \equiv \,
			\left( \begin{array}{cc}
			g_{11} & g_{12} \\
			g_{21} & g_{22}
			\end{array}
			\right).
\eeq

We have used a generic symbol $g_{ij}$ to denote the blocks for
both time one and two. It must be emphasized that these are {\it
not the same}. We just do not want to clutter up the symbols
with the dimensionality of the blocks, and we hope that this
does not create any undue confusion. Here we have  assumed that
$N$ is divisible by eight.
One may note that the structure of the matrices is a reflection
of the structures of the quantum baker's map [1]. The
Fourier blocks are at classical period two points of the four bakers'
map.  The preservation of classical symmetries makes $B^{(2)}$
 a {\it real } quantum map too.

In figure 2 we show the ratios Tr $(B^{2})/$Tr $(B^{(2)})$,
Tr $(B^{\prime 2})/$Tr $(B^{(2)\prime})$, corresponding to
the quantum four bakers' map and the quantum baker's map.  In the last ratio
the plot is split into the ratio of the real part of the traces
and the imaginary part of the traces. In the first ratio the
operators are themselves real, and so will be the traces.
The bold line in the figure represents the case of the four bakers'
map, while the dashed lines correspond to the baker's map.
The convergence in the case of the four bakers' map of the traces
of the quantum and the semiquantum propagators is apparent.
 This is to be compared with the
behaviour of the quantum baker's map, where the deviations of
the ratios from unity are quite large. This is in spite of the
fact that at the same value of the Planck constant, the {\it effective}
Planck constant for a single baker of the four bakers' map
is four times larger. Thus a more comprehensive study of the
quantitative differences between the semiquantum and quantum
propagators of the four bakers' map is warranted.

\subsection{Semiclassics of Semiquantum at Time Two}

\hspace{.5cm} We can provide a semiclassical analysis of the time two
semiquantum propagator with relative ease, and we do so here.
We wish to once more decompose the dynamics
according to classical symbol sequences. There are four independent
bakers and hence we can write the dynamics as a full shift on four
independent binary symbols. We will concentrate on two of them
as the other two will be related to these by $R$ symmetry.
Explicitly we will denote by $b_{00}$ the matrix
\beq
	e^{i \pi N/8}\,G_{N}^{-1} ( \delta_{11} \otimes g_{11})
\eeq
Here $\delta_{ij}, \; i,j=1,\ldots 8$ are  eight dimensional
matrices with a one at the i$^{\mbox{th}}$ row and the j$^{\mbox{th}}$
column, the rest being zero. They
serve to indicate the positioning of the $N/8$ dimensional blocks
in the mixed representation of $B^{(2)}$. Compare with eq.(35). The above
equation
tells us to select only the upper left most block in the
mixed representation. The trace of this operator would be
semiclassically governed by the fixed point (0,0).
we write similarly
\beq
	\ol{b}_{00} \;=\; e^{i \pi N/8} \,G^{-1}_{N} (\delta_{81} \otimes g_{21})
\eeq
\beq
	b_{10} \;=\; G^{-1}_{N} (\delta_{32} \otimes g_{11})
\eeq
\beq
	\ol{b}_{10} \;=\; G^{-1}_{N} (\delta_{62} \otimes g_{21})
\eeq
The overbars are only to indicate the blocks that are classically reflections
of each other.
As for time one we may now invoke both $R$ and $R\;T$ symmetries to write
the complete trace as
\beq
	\mbox{Tr} \;B^{(2)} \;=\; 4 \;\mbox{Tr} \;(b_{00}+\ol{b}_{00}
	+b_{10}+\ol{b}_{10}).
\eeq

An explicit calculation gives us
\beqa
&&\mbox{Tr}\;( b_{00}+\ol{b}_{00}) \;=\; \nonumber \\
&& e^{i \pi N/8} \, \frac{2}{N}\;\sum_{m=0}^{N/8-1}\, \sum_{n=0}^{N/8-1} \;
e^{-6 \pi i (m+1/2)(n+1/2)/N}\;(1\,+\, e^{3 \pi i (m+1/2)/4})
 \\
&&=\;  e^{i \pi N/8} \, \frac{4}{N} \; \mbox{Re} \left[ \,\sum_{m=0}^{N/8-1}\,
\sum_{n=0}^{N/8-1}
\left(e^{-6 \pi i (m+1/4)(n+1/2)/N} \right) \right] \\
&&=\;e^{i \pi N/8}\,  \frac{2}{N} \sum_{m=0}^{N/8-1} \;
\frac{\sin \left(3\pi(\,m +1/2)/4\right)}
{\sin \left(3 \pi(\,m+1/2)/N \right)}
\eeqa
The corresponding double sum that led to eq.(25) for the usual
baker's map is [1]
\beq
\frac{2}{N}\; \sum_{m=0}^{N/4-1}\, \sum_{n=0}^{N/4-1}\; e^{-6 \pi i
(m+1/2)(n+1/2)/N}.
\eeq
 The regular behaviour of the real part of the trace of the
usual quantum baker's map  was already noted in [1],
and shown that in fact the real part was in the ``regular regime''.
 The four bakers' map picks out as relevant only the
real part of the trace, eq.(43), thus it should not come as a
surprise if there are no singular traces in the model. This is
also expected since there are {\it no} corners on the torus
defining the phase space of the four bakers' map. Thus there are
no ``half hyperbolic points''. However the deeper reasons, if
any, for the regular behaviour of the real part is not known to
the author, and the four bakers' map may provide an interesting
model to shed some light on this question as it chooses only the
regular real part.

As was done for time one, in eq.(43) reflect the summation lattice
about $m=0$ and $n=0$,
to obtain the following.
\beq
\mbox{Tr}\, (b_{00}\,+\,\ol{b}_{00})\;=\;
e^{i \pi N/8} \, \frac{1}{N} \sum_{m=-N/8}^{N/8-1}\; \sum_{n=-N/8}^{N/8}\;
e^{-6 \pi i (m+1/2)(n+1/2)/N}
\eeq
which is a sum in the regular regime, there are no corners in
the summation, and it is again legal to replace the sum asymptotically
by an unbounded continuum approximation, resulting in
\beq
\mbox{Tr}\, (b_{00}\,+\, \ol{b}_{00})\; \sim \;
e^{i \pi N/8} \, N\; \int_{-\infty}^{\infty} dp \,\int_{-\infty}^{\infty}
 dq \; e^{-6 \pi i N pq} \;=\; \frac{1}{3}\, e^{i \pi N/8}.
\eeq
More elaborate, ``brute force'' [1]  methods of evaluating the semiclassical
trace confirm the above result, thus we are justified in
adopting conventional wisdom in this model.

We can write the equations for the ``regular regime'' by similar explicit
evaluations. In particular we write,
\beq
\mbox{Tr} \, b_{10}\; =\; \frac{2}{N}\;e^{i \pi N/3} \; \sum_{m=N/8}^
{N/4-1} \;
\sum_{n=N/4}^{3N/8-1}\; e^{-6 \pi i (n+1/2 -N/3)(m+1/2-N/6)/N}
\eeq
and
\beq
\mbox{Tr} \, \ol{b}_{10}\; =\;\frac{2}{N} \;  e^{-i \pi N/3} \; \sum_
{m=N/8}^{N/4-1} \;
\sum_{n=5N/8}^{3N/4-1}\; e^{-6 \pi i (n+1/2 -2N/3)(m+1/2-N/6)/N}
\eeq
In eq.(49) when we  reflect  the $n$ summation by means
of the transform $n \, \ra \, N-n-1$, we get the exact conjugate
of $\mbox{Tr} \, b_{10}$.
Thus we get, after applying the unbounded continuum
approximation,
\beq
\mbox{Tr} \,( b_{10}\, + \ol{b}_{10})  \; \sim \;4\;N\;
\mbox{Re} \left[
\, e^{i \pi N/3}\;  \; \int_{-\infty}^
{\infty} dp \int_{-\infty}^{\infty} dq \; e^{-6 \pi i N (q-1/3)(p-1/6)}
 \right] \eeq
\beq
= \frac{4}{3} \;\cos \left( \frac{\pi N}{3} \right)
\eeq
and thus finally the complete trace of $B^{(2)}$ can be written
from eq.(41)  as
\beq
\mbox{Tr} \, B^{(2)} \; \sim \; e^{i \pi N/8} \, \frac{4}{3} \, +\,
\frac{16}{3}\,
\cos \left( \frac{\pi N}{3} \right).
\eeq
Comparing these results with the corresponding formulae for the usual
quantum baker's map eqs.(25,26) we see that once more the four baker traces
are semiclassically four times the {\it regular} part when 16
divides $N$. We note here, without proof, that while the traces of
the quantum and the semiquantum propagators
had just enough cancellations to remove terms of order $\log \,
(\hbar)$, the  Euclidean norm marginally diverges whether the
classical map has corners (usual baker's map) or not (four
bakers' map). Further work on this is needed as the results are
inconclusive.

\section{ Periodic Orbit Sum }

\hspace{.5cm} Following the results of the previous section it does not
require great intuition to understand the semiclassical periodic
orbit sum of the four bakers' map. In this section we extend the
semiclassical analysis of the semiquantum maps upto times where
it is defined, and thereby arrive at a periodic orbit sum for
the quantum four bakers' map. The existence of such a sum for
the usual quantum baker's map is well known [1,3],
but the four bakers' map has the added advantage that there are
no anomalous terms and we do not have to treat the fixed points
specially. The symbol $T$ has been used in other sections of
this article to denote translational symmetry, but in this section
it will denote time.

We have to first describe the time $T$ semiquantum propagator.
We will concentrate first on the upper left half of the
propagator in the {\it mixed } momentum-position representation.
Thus the blocks populating this area will be $N/2^{T+1}$
dimensional matrices which we will write generically as $g_{11}$.
The construction of this propagator requires that $2^{T+1}\,|\,N$; we
will assume a slightly stronger condition that $2^{T+2}\,|\,N$.
This assumption will do away with any phases that globally sit
on blocks (i.e., they become one). The semiquantum propagator is
defined only upto time $T+1$, and the expressions we write down
will be valid only upto time $T$. We locate the $g_{11}$ blocks,
like the blocks are located for the usual quantum baker's map,
at the classical periodic orbits. The upper left hand corner of the
matrix corresponds to the origin of the classical phase space.

Let $\nu_{L}$ be a binary string of length $L$ and its binary
evaluated value be $\nu$. We denote by $\ol{\nu}$ the bit reversed,
binary evaluated integer of the same string. For example, if
$L=2$ and $\nu_{L}=01$, then $\ol{\nu_{L}}=10$, $\nu=1$ and $\ol
{\nu} = 2$.   $(\nu,\, {\ol{\nu}})$
uniquely determines a classical periodic orbit of length $L$ and
a $g_{11}$ block in the semiquantum propagator. The classical periodic
orbit is located [1,3]  at
\beq
 q_{0}\,=\, \frac{\nu}{2(2^{L}-1)}, \;\;\;p_{0}\,=\,
\frac{\ol{\nu}}{2(2^{L}-1)}.
\eeq
This refers to the periodic orbits in the lower left, principal baker's
map. It differs from the location of the periodic orbits in the
usual baker's map by a trivial scaling factor of $1/2$.
The semiquantum propagator block that is controlled by this
periodic orbit is a matrix of dimension $N/2^{T+1} \, \times \, N/2^{T+1}$
and is given in the mixed representation by
\beq
\sqrt{\frac{2^{T}}{N}} \,\exp \left[ - \frac{2^{T+1} \pi i}{N}
 (m+\frac{1}{2} -\frac{\nu N}
{2^{T+1}})(
n+\frac{1}{2} -\frac{\ol{\nu}N}{2^{T+1}}) \right]
\eeq
where
\[
m\,=\, \nu N/2^{T+1}, \ldots,(\nu +1)N/2^{T+1} -1,\;\;
\]
\[
n\,=\, \ol{\nu} N/2^{T+1}, \ldots,(\ol{\nu} +1)N/2^{T+1} -1.
\]

For convenience we define
\[ M \, \equiv \, N/2^{T+1}. \]
The ``conjugate block'' corresponds to the classical fixed point
at $( q_{0}, \, 1\,-\, p_{0})$, and exists in the upper
right half baker of the classical phase space. The semiquantum blocks
generically denoted as $g_{21}$,  controlled by such  periodic points,
populate the lower left half of the propagator in the mixed representation.
If the reader is confused, she is advised to turn to the special
cases of time one and two in the previous section for comparisons.
This block is once more of linear dimension $N/2^{T+1}$, and is
given by
\beq
\sqrt{\frac{2^{T}}{N}} \,\exp \left[  -\f{2^{T+1} \pi i}{N} (m+\f{1}{2} -
\f{(2^{T+1}-\nu-1) N}{2^{T+1}})(
n+\f{1}{2} -\f{\ol{\nu}N}{2^{T+1}})\right]
\eeq
where
\[
m\,=\, (2^{T+1}-\nu-1) N/2^{T+1}, \ldots,(2^{T+1}-\nu )N/2^{T+1} -1,\;\;
\]
\[
n\,=\, \ol{\nu} N/2^{T+1}, \ldots,(\ol{\nu} +1)N/2^{T+1} -1.
\]

After a Fourier transform to the pure position representation we
trace the correponding operators to get
\[
\mbox{Tr}\, b_{\nu}\,=\, \frac{\sqrt{2^{T}}}{N}\, \sum_{n=\nu M}
^{(\nu +1)M-1}\; \sum_{m=\ol{\nu}M }^{
(\ol{\nu} +1)M-1}\; \exp \left[ \frac{2 \pi i}{N} (m+\f{1}{2}
)(n+\f{1}{2}) \right]
\cdot\,
\]
\beq
\exp \left[ -\f{2^{T+1} \pi i}{N} (n+\f{1}{2} -\f{\nu N}{2^{T+1}})(
m+\f{1}{2} -\f{\ol{\nu}N}{2^{T+1}})\right].
\eeq
and
\[
\mbox{Tr}\, \ol{b_{\nu}}\,=\, \frac{\sqrt{2^{T}}}{N}\,
\sum_{n=(2^{T+1}-\nu-1) M}^{
(2^{T+1}-\nu)M-1}\; \sum_{m=\ol{\nu} M}^{
(\ol{\nu} +1)M-1}\; \exp \left[ \f{2 \pi i}{N} (m+\f{1}{2})
(n+\f{1}{2})\right]\,
\cdot
\]
\beq
\exp \left[ -\f{2^{T+1} \pi i}{N} (n+\f{1}{2} -\f{(2^{T+1}-\nu-2) N}{2^{T+1}})(
m+\f{1}{2} -\f{\ol{\nu}N}{2^{T+1}})\right].
\eeq

We can write the sum in eq.(56) as
\[
\mbox{Tr}\, b_{\nu}\,=\, \frac{\sqrt{2^{T}}}{N}\,
\, \exp \left( \f{i \pi N \nu \ol{\nu}}{2(2^{T}-1)}\right) \cdot
\sum_{n=\nu M}
^{(\nu +1)M-1}\; \sum_{m=\ol{\nu} M}^{
(\ol{\nu} +1)M-1}\; \cdot
\]
\beq
 \exp \left[ -\f{2 \pi i}{N}
(2^{T}-1)(n+\f{1}{2}-\f{\nu N}{2(2^{T}-1)})(m+\f{1}{2}- \f{\ol{\nu}N}{2(2^{T
}-1)}) \right].
\eeq
Similarly from eq.(57) we get
\[
\mbox{Tr}\, \ol{b}_{\nu}\,=\, \frac{\sqrt{2^{T}}}{N}\,
\, \exp \left( -\frac{i \pi N \nu \ol{\nu}}{2(2^{T}-1)}\right)
 \cdot \sum_{n=(2^{T+1}-\nu-1) M}^{
(2^{T+1}-\nu )M-1}\; \sum_{m=\ol{\nu} M}^{
(\ol{\nu} +1)M-1}\; \cdot
\]
\beq
 \exp \left[ -\f{2 \pi i}{N}
(2^{T}-1)(n+\f{1}{2}+\f{\nu N}{2(2^{T}-1)}-N)(m+\f{1}{2}-
\f{\ol{\nu}N}{2(2^{T}-1)})\right].
\eeq

In the last equation if we reflect the $n$ summation about the
center, $(n\, \ra \, N-n-1)$, we get the  complex conjugate of the other trace,
i.e.,
\beq
	\mbox{Tr} \,b_{\nu} \;=\; (\mbox{Tr} \, \ol{b}_{\nu})^{\ast},
\eeq
where the $\ast$ indicates complex conjugation. Thus we now have
\beq
\mbox{Tr} \, (b_{\nu} \, +\, \ol{b}_{\nu})\;=\; 2 \, \mbox{Re}
\, ( \mbox{Tr} \, (b_{\nu})),
\eeq
which is in the regular regime irrespective of whether the fixed
point is at the corner or not (see the discussions in the
previous section). However at the corner, that is when $\nu$ is a
pure string of 1 or 0, there is an additional step of reflecting the
summation lattice about $m=0$ and $n=0$ before we will be able
to apply the unbounded continuum approximation. This step leads
to a factor of a quarter, which we will account for below.
With this mind, we can replace the
summations in eq.(58) by integrals and do a stationary phase
approximation.
 We note that the relevant periodic point where the
phase is stationary is $(q_{0},p_{0})$, and is given by eq.(53).

Thus we get $\mbox{Tr}(b_{\nu}\,+\, \ol{b}_{\nu}) \; \sim \;$
\[
2 \sqrt{2^{T}}\; N \; \mbox{Re} \left[ \exp \left( \f{i \pi \nu \ol
{\nu}N}{2(2^{T}-1)} \right)
 \int_{-\infty}^{\infty} dq \; dp \;  \exp \left( -2 \pi i N(2^{T}-1)
(q-q_{0})(p-p_{0}) \right) \right]
\]
\beq
= \frac{ \cos  \left(\frac{ \ts \pi \nu \ol{\nu}N}{ \ts 2(2^{T}-1)}\right)}
{\sinh
\left(  \frac{ \ts T}{ \ts 2} \, \ts \log 2  \right)}.
\eeq
The complete trace of the semiquantum operator at time $T$ is,
due to  $R$ symmetry,
\[
\mbox{Tr}\,  B^{(T)}\;=\; 2 \,\sum_{L(\nu)=T}\; \mbox{Tr} \, (b_{\nu} \,+\,
 \ol{b}_{\nu}).
\]
The summation extends over all possible binary strings of length
$T$.

The fixed points at the center (1/2,1/2)  and corner (0,0) of
the classical phase space  have
to be treated with more care, as has been stated above. The
extra  factor
of one quarter for these points should hardly come as a surprise,
as a reflection about the lines $q=1/2$, $p=1/2$, do not produce
any new periodic points, unlike the ``bulk'' periodic orbits which
create three more copies of themselves. In other words while
the fixed points at (1/2,1/2) and the four corners belong to
{\it all} four of the bakers, the other periodic points belong
to individual bakers and thus there are four copies of each.
  This can be accounted
for by calculating the fixed point contributions explicitly.
Thus we write the final semiclassics of the four bakers' map which
we identify with the semiclassics of the semiquantum map at time
$T$ as,
\beq
\mbox{Tr} \, B^{T} \; \sim \; \mbox{Tr} \, B^{(T)} \; \sim \;
\frac{1}{\sinh \left( \frac{T}{2} \log 2 \right)} \, \left( 1\,
+\, 2 \, \sum_{L(\nu)=T}^{\hspace{.25in}\vspace{-.1in} \prime} \, \cos (2 \pi N
S_{\nu}) \right).
\eeq
Here $S_{\nu} \; =\; \f{ \ts \nu \ol{\nu}}{\ts 4(2^{T}-1)}$ is the
action for the classical periodic orbit. It differs by a trivial
factor of a quarter from the action for the usual baker's map [3].
The prime indicates that the sum is taken over all binary strings
of length $T$ except the ones which have exclusively 1 or
exclusively 0 in them. The contributions from such orbits have
been already explicitly added. The formulae of the previous
section are special cases of this formula, for $T=1,2$.

This formula is then in the form of a Gutzwiller-Tabor periodic
orbit sum, which expresses the trace of the propagator as an
approximate sum over classical periodic orbits. It is
essentially the same as the real part of the periodic orbit sum
for the usual quantum baker's map, with the important
modification that there will be no anomalous terms in the trace
of the exact propagator. In figure 3 we compare the periodic
orbit sum with the exact quantum traces in the time domain.
Note that the short periodic orbit traces are very good approximations,
unlike in the case of the usual baker's map, where the effect of
the fixed points non-generic character produced large errors.

The energy domain is simply a Fourier transform away, and the
reliability of eigenvalues estimated depends crucially on the
time domain input. The value of $N$, the inverse of the Planck
constant is 1024. We see that the
periodic orbit sum {\it tracks} the exact trace for a substantially
long time, as compared to the log time for each individual
baker's map. Although the periodic orbit sum was derived under
the assumption that $2^{T+2}|N$, we can evaluate the sums beyond
such times.
We have compared the traces for twenty time steps,
and we find the continuation of tracking. While this long time semiclassical
 accuracy has been
noted earlier [2], the individual periodic orbit traces can sometimes be a
very bad estimate of the exact trace. It might be that the error
is weakly bounded in time. We get a similar behaviour in another
quantum map on the torus, the sawtooth map [19].
Thus the initial hope that the quantum
baker's map will represent generic  behaviour of quantum non-integrable
systems is at least partially restored.

Recently there have been results that seem to improve the
accuracy of the periodic orbit formula, either by accounting for
orbits that are just about to be ``born'' out of a bifurcation [17]
or by calculating higher order corrections [18]. It would thus
be of interest to understand the possible ways in which the periodic
orbit sum for the baker's map can be improved. We believe that
with the removal of terms logarithmic in the Planck constant
from the semiclassical formulae  the way has been cleared for
more such detailed studies.

\section{Discussions}

\hspace{.5cm} The  rigorous dissection of the quantum baker's map
in ref. [1], showed up many non-generic features of the map,
and the principal questions that grew out of it was whether these
properties are stable against changing quantizations and  models.
We have demonstrated here that such features are not stable
against  modifications of quantization. We showed in particular that the
quantum bakers as social animals tend to have normal semiclassical
features.
While there still remains the question
of {\it why} the boundary conditions mattered so much in the
quantum baker's map, it is clear that  the generator of such discrepancies
 lies in the unusual configuration of the fixed point, whose
trace contribution was not that of a regular hyperbolic point.

The semiclassics of the quantum four bakers' map follows a
periodic orbit sum and there are no anomalous terms in the traces
of the propagators generating this sum. Thus it is an ideal
testing ground for the periodic orbit sum, and its validity in the
time and energy domains. It has been observed that the traces of the baker's
map are valid much beyond the $ \log$ time [1,2]. While the
propagators that were used to generate these traces, the
semiquantum propagators,  themselves
do not exist, evaluating the periodic orbit sum seems to produce
reasonably accurate traces. We can expect that the quantum four
baker's map will follow similar behaviour, as it is essentially very
similar to the usual baker's map. In fact we have seen that not
only classically is the four bakers' map redundant, but also
semiclassically the information from a single baker's map, the
fundamental domain,  was sufficient.
The genericity of the results reported here depend upon the
investigation of other quantum maps with chaotic classical limits.
The semiclassical theory of the quantum sawtooth map [19]
shows similar features.

 The four bakers' map may prove to be interesting for other
reasons, though it has now lost some simplicity that the
baker's map had. We intend to pursue its study elsewhere.
The novel feature of the four bakers' map is the tunnelling
that occurs between the classically isolated bakers. Such
tunnelling was found to be abnormally large in another
juxtaposed baker map [11]. The tunnelling occurs between
regions in phase space which have chaotic dynamics within them
and are separated by a separatrix. Coherent tunnelling effects,
which we have not discussed above,
have been observed between diagonally opposite bakers.
It is not clear at this stage
how relevant such tunnelling effects are for generic dynamical
systems.
 An exact symbolic decomposition of the quantum propagator
exists for the quantum baker's map [1], based on the
possible classical transitions at any given time.  The
possibility of tunnelling between the regions of the four
bakers' map makes such a symbolic decomposition more tedious,
in particular there will be classically forbidden symbol
sequences in the decomposition of the exact quantum propagator.

\vspace{1cm}
\begin{center}
{\bf Acknowledgements}
\end{center}
\hspace{.5cm}It is a pleasure to thank Professor N.L. Balazs and Dr. A. Voros
for many discussions and constant encouragement. I am grateful
for the hospitality of the Service de Physique Theorique, Saclay,
where the principal ideas of this note became clear. It is also a
pleasure to thank  Professor V.B. Sheorey for several discussions.

\newpage

\begin{center}
 {\large{ \bf References}}
\end{center}

\begin{enumerate}

\item {\sc M. Saraceno, and A. Voros}, {\sl Physica} {\bf 79 D} (1994), 206.

\item {\sc P.W.O' Connor, S. Tomsovic, and E.J. Heller}, {\sl Physica} {\bf
55D} (1992), 340.

\item {\sc A.M.O. De Almeida, and M. Saraceno}, {\sl Ann. Phys., (N.Y.)}
{\bf 210} (1991), 1.

\item {\sc N.L. Balazs, and A. Voros}, {\sl Ann. Phys., (N.Y.)}{\bf 190}
(1989), 1.

\item {\sc M. Saraceno}, {\sl Ann. Phys., (N.Y.)} {\bf 199} (1990), 37.

\item {\sc L.E. Reichl}, ``The Transition to Chaos in Conservative
Classical Systems: Quantum Manifestations'', Springer Verlag, New York,
1990.

\item{\sc P.W.O' Connor, and S.Tomsovic}, {\sl Ann. Phys.,
(N.Y.)}{\bf 207} (1991), 218.

\item{\sc J. Schwinger}, ``Quantum Kinematics and Dynamics'',
Benjamin, New York, 1970.

\item {\sc V.I. Arnold, and A. Avez}, ``Ergodic Problems of Classical
Mechanics'' WA Benjamin, New York (1968).

\item{\sc A. Lakshminarayan, and N. L. Balazs}, {\sl J. Stat. Phys.} {\bf 77}
(1994) 311.

\item{\sc A. Lakshminarayan}, Doctoral Dissertation, Dept. of
Physics, S.U.N.Y at Stony Brook (N.Y., 1993).

 \item {\sc M.Abramowitz, and I.A.Stegun},``Handbook of mathematical Functions
,'', U.S. National Bureau of standards, Washington. D.C., 1964.

\item{\sc I.C. Percival, and F. Vivaldi}, {\sl Physica} {\bf 30D}
(1988), 164.

\item {\sc J.H. Hannay, and M.V. Berry}, {\sl Physica} {\bf 1D} (1980), 267.

\item{\sc J. P. Keating}, {\sl Nonlinearity} {\bf 4} (1991), 335.

\item{\sc J. Ford, G. Mantica, and G.H. Ristow}, {\sl Physica}
{\bf 50D} (1991), 493.

\item{\sc M. Kus, F. Haake, and D. Delande}, {\sl Phys. Rev. Lett.} {\bf 71}
(1993), 2167.

\item{\sc P. Gaspard, and D. Alonso}, {\sl Phys. Rev.} {\bf
A47} (1993), R3468.

\item{\sc A. Lakshminarayan}, {\sl Phys. Lett.} {\bf 192 A} (1994), 345.

\item{\sc M.C. Gutzwiller}, ``Chaos in Classical and Quantum Mechanics'',
 Springer (New York, 1990).

\item{\sc M.A. Sepulveda, S.Tomsovic and E.J. Heller}, {\sl Phys.
Rev. Lett.} {\bf 69} (1992), 402.
\end{enumerate}

\newpage
\begin{center}
{\large \bf Figure Captions}
\end{center}

\noindent {\bf Figure 1.} The four bakers' map before and after a
transformation.
Shown are the principal partitions of the map. The fundamental
baker comprises of the partitions 1 and 5. The rest are obtained
by reflections.

\noindent {\bf Figure 2.} Comparison of the traces of the quantum and the
semiquantum operators at time two  for the baker's and the four bakers' maps.
A represents the case of the four bakers' map, which is a real
map. B and C represent the ratio of the real and imaginary parts
of the traces for the case of the usual baker's map.

\noindent {\bf Figure 3.} Comparison of the periodic orbit sum traces
with the exact traces, for the case when $N=1024.$ The first solid
line in the ``doublets''  represents the periodic orbit sum, while the
artificially shifted lines represent the quantum traces.

\newpage
\begin{center}
{\large \bf Appendix }
\end{center}

\hspace{.5cm} The original formulation of the quantum four bakers' map was in
the
form of eq.(14), while much of the later results depend upon the
matrix form of eq.(19). We derive here the latter from the former.
\[
B\;=\; G_{N}^{-1} \, ( -T \, u_{0} \, T \, + \, u_{0} ),
\]
where
\[ T \,=\, e^{i \pi N /4} \, t_{1} \, t_{2}
\]
with
\[ t_{1} \, =\, U^{N/2}, \; \; \mbox{and} \;\;t_{2}\,=\,
V^{-N/2}. \]
$U$ and $V$ are translation operators, in momentum and position,
 on the torus and have been
defined in eqs.(5,6). We note that $t_{1}$ and $t_{2}$ commute
because we have assumed $N$ to be divisible by 4. The position
representation of $V^{-N/2}$ is given by a matrix which we conveniently
write as
\beq
t_{2} \;=\; \left( \begin{array}{cccc}
	0 &0 & -I_{N/4} & 0 \\
	0 & 0& 0& -I_{N/4}\\
	I_{N/4}&0&0&0\\
	0& I_{N/4}&0&0
	\end{array}
	\right),
\eeq
where $I_{N/4}$ are $N/4$ dimensional identity matrices.
Then using the matrix representation of $u_{0}$, (11), it is
easy to see that
\beq
u_{1} \, \equiv \, t_{2} \, u_{0} \, t_{2} \, =\, \left( \begin{array} {cccc}
	-g_{22} & 0 & 0 & g_{21} \\
	0&0&0&0\\
	0&0&0&0\\
	g_{12}&0&0&-g_{11}
	\end{array}
	\right).
\eeq

Now we separate out diagonal $t_{1}$ into four $N/4$ dimensional
diagonal blocks which we write as $ e^{i \pi/2}\, a_{j}$, with
$j=1,\ldots,4$. $(a_{1})_{mn} \,=\, (-1)^{m} \, \delta_{mn}, \,
m,n =0, \ldots, N/4-1$ and $(a_{j})_{mn} \, =\,(-1)^{N(j-1)/4} a_{1}$,
  where $j=1,2,3,4$. We then get
\[
-T\, u_{0} \, T \, + u_{0}\, =\, -t_{2}\, u_{1}\,
t_{2}\,+u_{0}\,  =\,
\]
\beq
\left( \begin{array}{cccc}
	a_{1} g_{22} a_{1} & 0 & 0 & -a_{1} g_{21} a_{4}\\
	0 & g_{11} &g_{12} & 0 \\
	0 & g_{21}& g_{22} &0\\
	-a_{4}g_{12}a_{1} &0&0&a_{4} g_{11}a_{4}
	\end{array}
	\right).
\eeq
Now a simple explicit calculation yields
\beq \begin{array}{ll}
a_{1}g_{22}a_{1} \, =\, e^{i \pi N/4} g_{11},\,\,
 -a_{1}g_{21}a_{4} \, =\, e^{i \pi N/4} g_{12},\\
-a_{4}g_{12}a_{1} \, =\, e^{i \pi N/4} g_{21}, \, \,
a_{4}g_{11}a_{4} \, =\, e^{i \pi N/4} g_{22},
\end{array}
\eeq
which establishes the matrix form in (19).
A simple extension of the arguments here also give us the matrix
form of the semiquantum propagator at time two, eq.(35), besides showing
us that if $2^{T+2} \, |\, N$ then there are no additional phase
factors. This has been  have used in section 5 where we have not
included any global phase factors.

\end{document}